\documentclass[journal=jpcafh,manuscript=article,layout=twocolumn]{achemso}

\usepackage[utf8]{inputenc}
\usepackage{textcomp}
\usepackage[flushleft]{threeparttable}
\usepackage{gensymb}
\usepackage[normalem]{ulem}
\usepackage{chemformula}
\let\ce\ch
\usepackage{dcolumn}
\usepackage{url}

\makeatletter
\let\l@addto@macro\relax
\makeatother
\usepackage[fontsize=11pt]{scrextend}

\DeclareSymbolFont{matha}{OML}{txmi}{m}{it}

\title{Collisional excitation of \ch{HCN} by \ch{CO} to refine the modeling of cometary comae}

\author{Francesca Tonolo}
\affiliation[rennes]{Univ. Rennes, CNRS, IPR (Institut de Physique de Rennes), UMR 6251, Rennes, F-35000, France} 
\email{francesca.tonolo.1@univ-rennes.fr}

\author{Ernesto Quintas-Sánchez}
\affiliation[missouri]{Department of Chemistry, Missouri University of Science and Technology, Rolla, Missouri 65409, USA} 

\author{Adrian Batista-Planas}
\affiliation[missouri]{Department of Chemistry, Missouri University of Science and Technology, Rolla, Missouri 65409, USA} 

\author{Richard Dawes}
\affiliation[missouri]{Department of Chemistry, Missouri University of Science and Technology, Rolla, Missouri 65409, USA} 

\author{François Lique}
\affiliation[rennes]{Univ. Rennes, CNRS, IPR (Institut de Physique de Rennes), UMR 6251, Rennes, F-35000, France} 
\email{francois.lique@univ-rennes.fr}


\begin{document}

\maketitle

\begin{abstract}\label{sec:abstract}
We present the first dataset of collisional (de)-excitation rate coefficients of \ch{HCN} induced by \ch{CO}, one of the main perturbing gases in cometary atmospheres. The dataset spans the temperature range of 5--50~K. It includes both state-to-state rate coefficients involving the lowest ten and nine rotational levels of \ch{HCN} and \ch{CO}, respectively, and the so-called ``thermalized" rate coefficients over the rotational population of \ch{CO} at each kinetic temperature. 
The derivation of these coefficients exploited the good performance of the statistical adiabatic channel model (SACM) on top of an accurate interaction potential computed at the CCSD(T)-F12b/CBS level of theory. The reliability of the SACM approach was validated by comparison with full quantum calculations restricted at the lowest total angular momentum of the system. 
These results provide essential input to accurately model the distribution among the rotational energy levels and the abundance of \ch{HCN} in cometary atmospheres, accounting for deviations from local thermodynamic equilibrium that typically occurs in such environments. 
\end{abstract}


\section{Introduction}\label{sec:intro}
Cometary ices are among the most pristine materials of the Solar System, as they preserve the chemical conditions that existed in its earliest stages \cite{mumma2011chemical}. Thanks to recent advances in observational techniques, we can now observe with very high spatial and spectral resolutions the molecular content of cometary comae, $i.e.$, the ensemble of gases which sublimate from cometary ices by solar radiation, forming an atmosphere around the cometary nucleus \cite{bockelee2008cometary,roth2021leveraging,biver2022observations,cordiner2023gas}. However, the accurate interpretation of these spectra requires sophisticated radiative transfer models. A key challenge in this regard is describing how a molecule's energy is distributed among its rotational levels. At the typical low densities of cometary comae ($1<n< 10^{-10}$\,cm$^{-3}$), this distribution can strongly deviate from the local thermodynamic equilibrium (LTE). In these cases, accurate datasets of collisional coefficients with the most abundant perturbing species are crucial (as an example, the reader is referred to \citet{loreau2022effect}), and usually rely on state-of-the-art theoretical calculations. 

However, when dealing with heavy projectiles such as \ch{H2O}, \ch{CO} and \ch{CO2} (the most abundant gases in cometary comae), collisional calculations notably complicate: the high densities of states of the projectiles dramatically increase the number of channels to be considered, which makes full quantum scattering calculations unaffordable in terms of computational time and memory requirements. To address this, the statistical adiabatic channel model (SACM) was recently tested as an alternative. Developed by \citet{quack1975complex} and subsequently refined by \citet{loreau2018efficient}, this approach can generate reliable sets of collisional coefficients in favorable cases at significantly reduced computational cost. In terms of accuracy, the SACM is particularly effective for systems that form a stable intermediate complex, often showing discrepancies of less than a factor 2 with respect to the exact close coupling (CC) calculations \cite{loreau2018efficient, loreau2018scattering, balancca2020inelastic, pirlot2025collisional}. For cometary applications, the SACM currently offers the most practical solution to overcome the prohibitive costs associated with heavy projectiles and to deal with the low temperatures ($\sim$\,5-50~K) that rule out the application of (quasi)classical methods \footnote{The mixed quantum/classical theory (MQCT) offers a promising compromise in this direction, but it currently remains insufficient to accurately characterize collisional processes at temperatures below 50\,K \cite{joy2024mixed,mandal2024mqct}.}. For the \ch{CS}--\ch{CO} collisional system, it was recently shown that it can produce state-to-state outcomes differing by less than an order of magnitude from full quantum calculations \cite{godard2025promising}. 

In order to refine the modeling of cometary comae, we address in this work the collisional excitation of \ch{HCN} by \ce{CO}. 
\ch{HCN} is one of the most abundant nitrogen-bearing molecules in cometary comae \cite{bockelee2011overview}. Detected in both millimeter and IR wavelengths, \ch{HCN} primarily originates from direct sublimation of ices on the comet nucleus. Hence, it serves as a valuable tracer of the thermal and chemical history of pre-cometary material \cite{drahus2010hcn, villanueva2013modeling}. Understanding the abundance of \ch{HCN} in comets compared to that in the interstellar medium provides insights into the early conditions of the Solar System and the possible delivery of prebiotic molecules to Earth \cite{matthews2008hydrogen,jung2013mechanisms}. Moreover, the abundance of \ch{HCN} relative to other nitrogen-bearing species ($e.g.$, \ch{NH3}, \ch{N2}) can indicate the temperature at which cometary ices formed and the extent and nature of the chemical process they underwent \cite{mumma1993comets,magee2002hydrogen, biver2022chemistry}.
However, the only collisional investigations of cometary interest regarding \ch{HCN} concern its excitation processes induced by \ch{H2O} \cite{dubernet2019first,zoltowski2025collisional}. These data are not sufficient to model cometary comae at intermediate and large heliocentric distances, where the solar radiation is not intense enough to sublimate the water from the nucleus. Distant cometary comae are instead generally dominated by \ch{CO} \cite{ootsubo2012akari,yang2021discovery, ejeta2025infrared}. These comets are particularly relevant for study of the composition of the early Solar System because they typically contain materials not previously altered by solar radiation.
Moreover, the low densities that characterize cometary comae lead to strong deviations from LTE, making accurate collisional rate coefficients essential for reliable radiative transfer modeling. 

This work aims to bridge this gap by providing the first dataset of collisional coefficients for the rotational (de)-excitation of \ch{HCN} by \ch{CO}. The methodology and results are presented as follows: the first Section describes the employed computational strategy. There, in the context of the Born-Oppenheimer approximation, the characterization of the interaction potential and the scattering calculations used to solve the nuclear Schrödinger equation are presented in two different subsections. The subsequent Section presents and discusses the results. Finally, in the last Section the main conclusions are drawn. 

\section{Computational Details}\label{sec:comp}

\begin{figure}[t]
 \includegraphics[width=1.0\columnwidth]{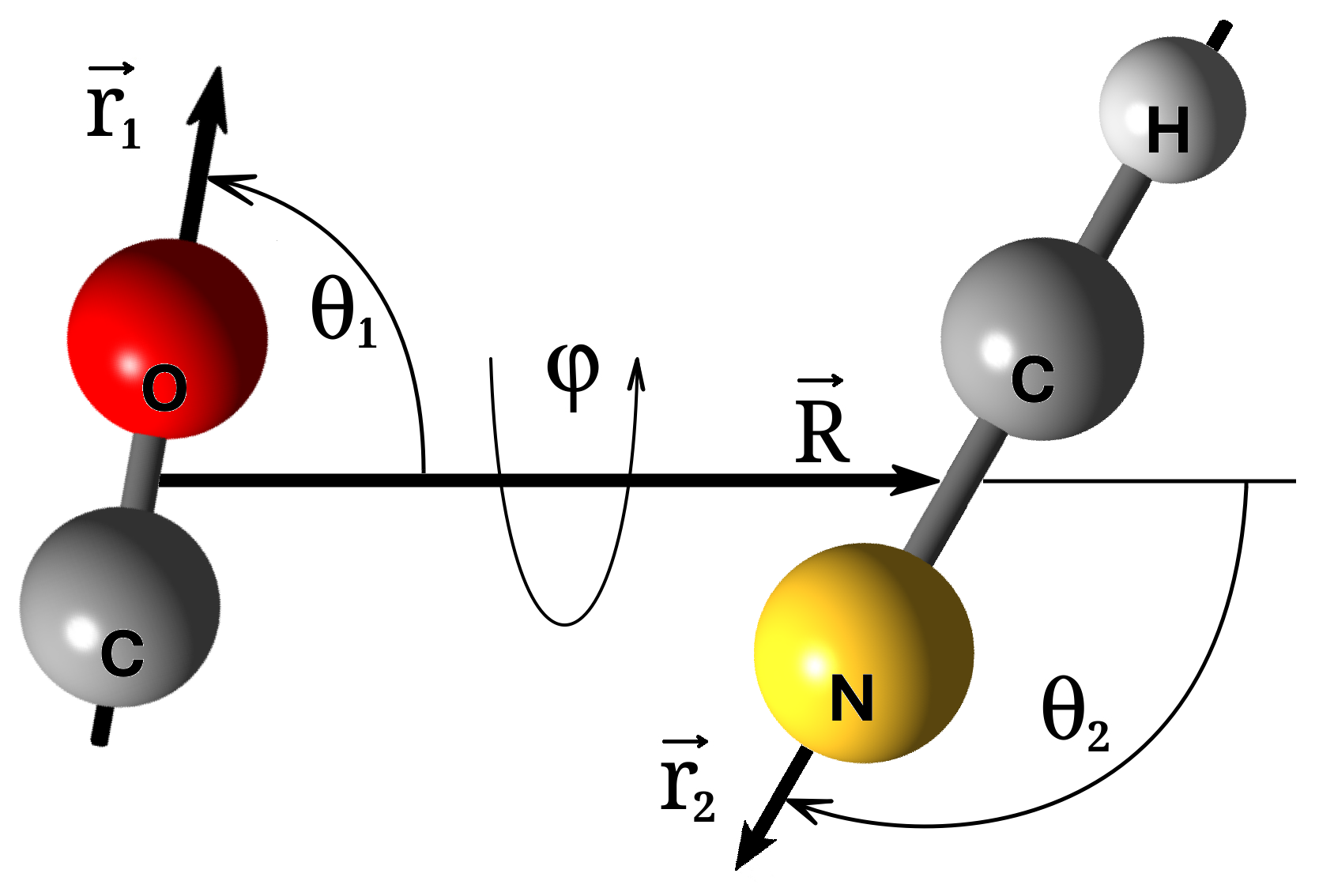}
 \caption{Coordinates used to describe the HCN--CO  interaction. 
  See the text for details.}
  \label{SR}
\end{figure}

\subsection{Potential Energy Surface}\label{subsec:PES}
The Jacobi coordinates used to represent the four-dimensional (4D) HCN--CO interactions are depicted in Figure~\ref{SR}. Note that for the representation of the PES, CO is monomer 1, and HCN is monomer 2. 
$\vec{R}$ is the vector between the centers of mass of the two fragments, and $\vec{r}_1$ and $\vec{r}_2$ are vectors aligned with each molecule.
Coordinate $R$ is the length of vector $\vec{R}$, while coordinates $\theta_1$ and $\theta_2$ represent (respectively) the angles between $\vec{R}$ and the vectors $\vec{r}_1$ and $\vec{r}_2$. 
The fourth coordinate is the dihedral (out of plane) torsional angle, labeled $\varphi$, which is the angle between the vectors $\vec{R}\times\vec{r}_1$ and $\vec{R}\times\vec{r}_2$. 

In line with previous works on other van der Waals (vdW) dimers composed of two linear molecules,\cite{castro2019computational,quintas2020computational,gancewski2021fully,Desrousseaux2021,quintas2021theoretical,zadrozny2022,ajili2022theoretical,olejnik2023ab,bostan2024mixed} the analytical representation of the PES was constructed using an automated interpolating moving least squares (IMLS) methodology, freely available as a software package under the name \textsc{autosurf}.\cite{quintas2018autosurf} 
The fitting basis and other aspects of the IMLS procedure  are similar to those employed for other previously treated systems that have been described in detail elsewhere.\cite{majumder2016automated,Dawes2018,quintas2018autosurf,quintas2021spectroscopy}
The shortest intermonomer center-of-mass distance considered is $R = 2.4$~{\AA}. 
The short-range part of the PES was restricted by excluding regions with repulsive energies above a maximum of  $6$~kcal/mol ($\sim2\,000$~cm$^{-1}$) relative to the separated monomers asymptote. The \textit{ab initio} data coverage in the fitted PES extended to $R=25$~{\AA}, while the zero of energy was set at infinite center-of-mass separation between the monomers.
To guide the placement of high-level \textit{ab initio} data, a lower-level guide surface was constructed using $3\,000$ points at the explicitly correlated CCSD(T)-F12a/VDZ-F12 level,\cite{Werner2011a} distributed using a Sobol sequence~\cite{sobol1976uniformly} biased to sample the short range region more densely. Since the two fragments are not identical, and each has only C$_{\infty v}$ symmetry, there is no additional symmetry to respect/exploit in the PES beyond that of the angle $\varphi$, for which energies were only computed in the reduced angular range: $0<\varphi<180^{\circ}$, as is the case in this treatment for all 4D systems.

\begin{figure}[t]
	\includegraphics[width=1.0\columnwidth]{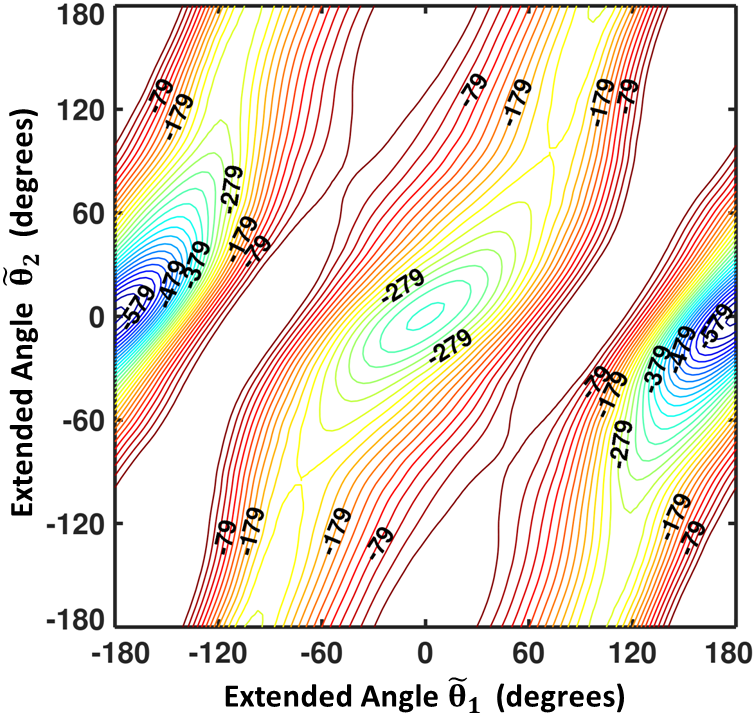}
	\caption{$R$-optimized contour plot of the PES as a function of the extended angles $\tilde{\theta}_1$ and $\tilde{\theta}_2$ for planar configurations ($\varphi=0^{\circ}$ and $\varphi=180^{\circ}$). For each pair of angles, the energy (given in cm$^{-1}$) is optimized with respect to the center-of-mass distance $R$. See text for details.}
	\label{Ropt}
\end{figure}

The final high-level PES was computed using the CCSD(T)-F12b method, extrapolated to the complete basis set (CBS) limit. 
The CBS extrapolation was performed using the VTZ-F12 and VQZ-F12 bases~\cite{peterson2008systematically} and the $l^{-3}$ formula.~\cite{feller2006sources}
All \textit{ab initio} calculations were performed using the Molpro electronic structure code package.\cite{werner2012-molpro} 

For construction of the PES, both monomers were held rigid. 
The bond distance for CO was fixed at $r_{\text CO}=1.128206$~{\AA}, the vibrationally-averaged bond distance for the ground ro-vibrational state of CO, consistent with its rotational constant of $B(\text{CO})=1.9225$\,cm$^{-1}$. \cite{winnewisser1997sub} 
For HCN, the equilibrium geometry is linear and the structural parameters employed in this study ($r_{\text HC} = 1.0655$~{\AA} and $r_{\text CN} = 2.2187$~{\AA}) are consistent with the experimental rotational constant, $B(\text{HCN})=1.4782$\,cm$^{-1}$. \cite{ahrens2002sub}
Masses of $15.9949146221$, $12$, $1.007825032$, and $14.0030740052$~$u$ were used for $^{16}$O, $^{12}$C, $^{1}$H and $^{14}$N, respectively.

To represent the long range (out to arbitrary distances), the PES switches smoothly to an analytic expression representing electrostatic, induction, and dispersion interactions between the fragments (truncated at 8th order). 
For the high-level PES, $3\,841$ points were used, resulting in an estimated global root-mean-squared fitting error of $0.8$~cm$^{-1}$ (excluding the long range where the fitting error is extremely small). The estimated error is $0.1$~cm$^{-1}$ for geometries corresponding to energies in the wells (below the asymptotic zero).
The analytical representation of the PES is available from the authors upon request.

\begin{figure}[t]
 \includegraphics[width=1.0\columnwidth]{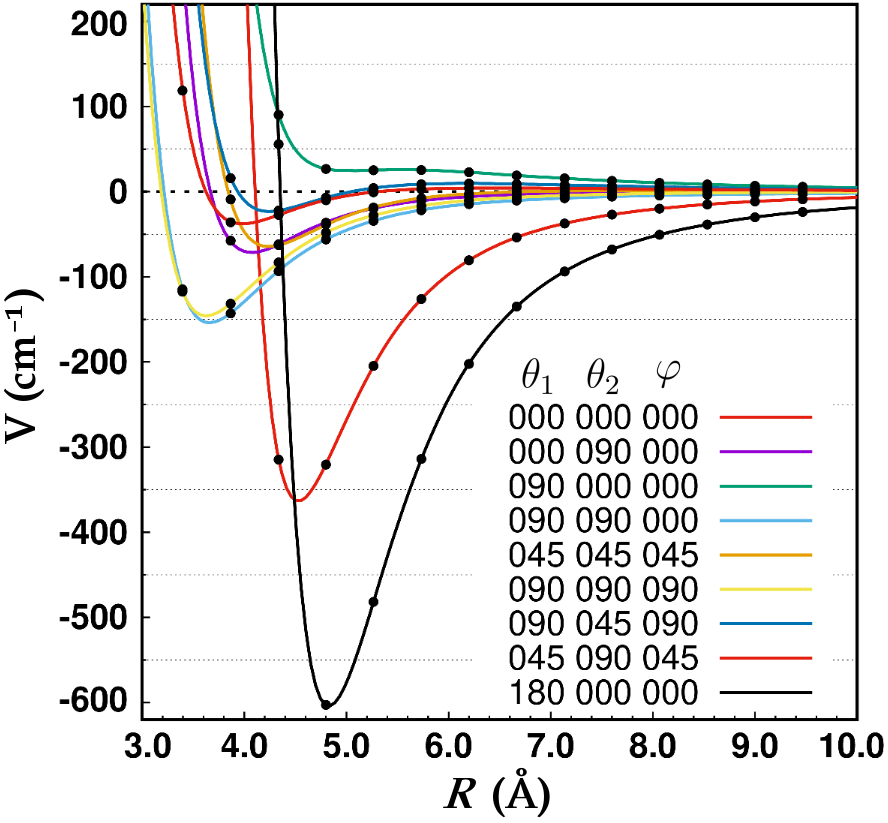}
 \caption{Various radial cuts defined by different orientational orientations of the monomers. In all cases,
energies are in cm$^{-1}$, lines represent the fitted PES, and points represent \textit{ab initio} calculations (not used in the fit).}
 \label{R-cut}
\end{figure}

Figure~\ref{R-cut} shows radial cuts through the PES for nine different angular orientations. The points represent \textit{ab initio} data (not included in the fit), while the lines plot the fitted PES. This helps one appreciate the anisotropy of the PES as reflected in the different values of $R$ at which the onset of a steep repulsive wall begins, as well as the accuracy of the fit.

\begin{table}[t]
\caption{\label{table1} Geometric parameters and  potential energy for stable structures in the PES. Energies are given relative to the asymptote. Units are \AA\,, degrees, and cm$^{-1}$.}
	\centering
    \begin{threeparttable}
	\begin{tabular}{lrr}
	               & GM \tnote{1}        & LM \tnote{1}        \\
	\hline
	$R$           & $4.826$   & $4.532$    \\
	$\theta_1$ & $180.0$   & $0.0$     \\
	$\theta_2$ & $0.0$    & $0.0$    \\
	$\varphi$        & ---     & ---      \\
	$V$           & $-604.13$ & $-363.32$  \\
    \hline
	\end{tabular}
      \begin{tablenotes}
    \item[1] {Defined in the text}.
  \end{tablenotes}
    \end{threeparttable}
\end{table} 

The PES is characterized by two minima, the global minimum (GM), a collinear arrangement where the C-atom of CO approaches the H-atom of HCN, and a local minimum (LM), also collinear, where the O-atom of CO approaches the H-atom of HCN. These minima are seen across the middle of Figure~\ref{Ropt}, with GM appearing twice in the plot due to the use of extended angles. The energies and structural parameters are given in Table~\ref{table1} where it is seen that GM ($E = -604.1$~cm$^{-1}$) is much more stable than LM ($E=-363.3$~cm$^{-1}$), although it is LM with a significantly shorter separation between monomers, due to the location of the CO monomer's center-of-mass.

\subsection{Scattering Calculations}\label{subsec:scatt}
The constructed state-of-the-art interaction potential was subsequently implemented in the scattering equations to solve the nuclear part of the Schrödinger equation and characterize the dynamics of the collisional system.

The most accurate method to solve scattering equations and to obtain the so-called scattering matrix, which contains all the collisional observables of the system, is the full quantum CC approach. However, for systems characterized by high densities of rotational states, combined with (a) deep well(s) of the intermediate complex (such as \ch{HCN} interacting with \ch{CO}), the number of coupled equations required for convergence becomes computationally prohibitive. 

To address this, as introduced above, we decided to adopt the SACM approach. This approximation requires to compute only the adiabatic curves that are obtained by excluding the nuclear kinetic term and diagonalizing the interaction Hamiltonian in a basis of rotational functions for the two colliders. This noticeably simplifies the scattering calculations, as each adiabatic curve is independent of the energy and hence needs to be computed only once for each value of $J_{\text{tot}}$. 
The calculations were performed by using the HIBRIDON code \cite{alexander2023hibridon}, which contains an implementation that permits one to retrieve only the adiabatic states. 
The scattering matrices were then rebuilt by including all the open adiabatic channels at each total energy, assuming that they occur with the same probability. This yields to state-to-state collisional cross sections, which can be subsequently averaged over a Maxwell–Boltzmann distribution of energies to obtain the corresponding rate coefficients.

As mentioned in the Introduction, the SACM method assumes that the intermediate complex formed during each collision lives long enough for energies to be statistically redistributed. This is a good approximation for the \ch{HCN} and \ch{CO} system, where the hydrogen bonding interaction with \ch{HCN} ($2.97$\,D) stabilizes the complex significantly. In fact, the well depth of the \ch{HCN}--\ch{CO} complex is nearly three times greater than that of \ch{CS}--\ch{CO}, for which the SACM has already shown good accuracy.\cite{godard2025promising} However, \ch{HCN} has a lower density of rotational states compared to \ch{CS}, and is therefore less efficiently stabilized in the potential well during collisions.
On the other hand, this characteristic reduces the computational demand and enables the exploration of a broader energy range.

The cross sections were calculated for transitions between rotational levels up to $j_2= 9$ for \ch{HCN} and $j_1=8$ for \ch{CO}, which are the most populated states up to 50\,K. At this temperature, the adiabatic channels associated with energy levels up to $\sim500$\,cm$^{-1}$ would typically be included in the calculations. In this energy range, the accessible rotational levels of \ch{HCN} and \ch{CO} are $j^{\text{max}}_2=18$ and $j^{\text{max}}_1=15$, respectively. However, the inclusion of the full rotational basis led to a substantial memory demand, reaching over 15000 open channels even at only $J_{\text{tot}}=2$. In order to ease the calculations from a computational point of view, the rotational basis was further truncated to $j_2^{\text{max}}=12$ and $j_1^{\text{max}}=10$. The accuracy of this truncation was ensured by confirming that the convergence of the rate coefficients at $J_{\text{tot}}=2$ is within 20\% with respect to the full rotational basis over the 5--50~K temperature range. Since the main contributions from higher values of $J_{\text{tot}}$ are expected to increase the centrifugal barrier for each collisional process, the resulting deviations can reasonably be considered within the uncertainty range of the fully converged calculations.

Using the reduced rotational basis, a partial wave expansion from $J_{\text{tot}}=0$ up to $J_{\text{tot}}=142$ was performed, ensuring convergence of the cross sections better than 5\%.
\begin{figure*}
\begin{center}  \includegraphics[scale=0.52]{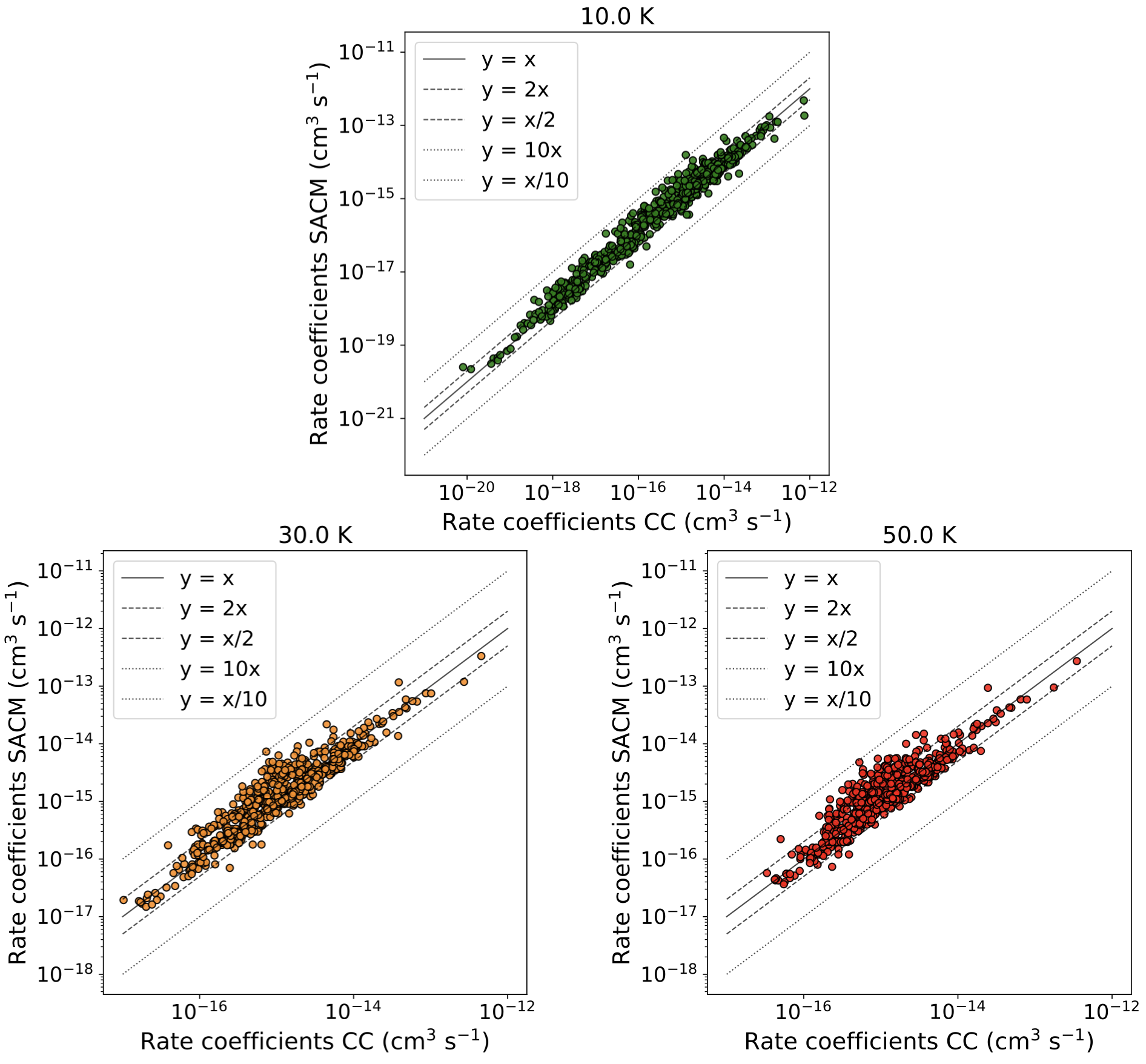}
\end{center}
\caption{Comparison between the state-to-state rate coefficients computed with the CC and SACM approaches at $J_{\text{tot}}=0$ for 10, 30 and 50\,K. }
\label{fig3}
\end{figure*}
Starting from the cross-sections for each $j_1j_2 \rightarrow j_1^{\prime}j_2^{\prime}$ transition ($\sigma_{j_1j_2 \rightarrow j_1^{\prime}j_2^{\prime}}$), the (de)-excitation rate coefficients ($k_{j_1j_2 \rightarrow j_1^{\prime}j_2^{\prime}}$) from 5 to 50\,K were derived by thermal averaging over the collisional energy ($E_c$):
\begin{equation}
\begin{aligned}
&k_{j_1j_2 \rightarrow j_1^{\prime}j_2^{\prime}}(T)= \left(\frac{8}{\pi \mu k_B^3 T^3}\right)^{1 / 2} \\
& \times \int_0^{\infty} \sigma_{j_1j_2 \rightarrow j_1^{\prime}j_2^{\prime}}\left(E_c\right) E_c \exp \left(-E_c / k_B T\right) \mathrm{d} E_c\,,
\end{aligned}
\end{equation}
where $k_B$ is the Boltzmann constant. 

In addition to the dataset of state-to-state collisional coefficients, thermalized rate coefficients for the (de)-excitation of \ch{HCN} were also retrieved. These coefficients were obtained by summing the $k_{j_1j_2 \rightarrow j_1^{\prime}j_2^{\prime}}(T)$ coefficients over all possible final states of \ch{CO} and averaging over the initial rotational states of \ch{CO}:
\begin{equation}
    \bar{k}_{j_2 \rightarrow j_2^{\prime}}(T)=\sum_{j_1} n_{j_1}(T) \sum_{j_1^{\prime}} k_{j_1j_2 \rightarrow j_1^{\prime} j_2^{\prime}}(T)\,,
\end{equation}
where $n_{j_1}(T)$ is the relative population of the projectile:
\begin{equation}
    n_{j_1}(T)=\frac{\left(2 j_1+1\right) \exp \left(-\frac{E_{j_1}}{k_{\mathrm{B}} T}\right)}{\sum_{j_1^{\prime}}\left(2 j_1^{\prime}+1\right) \exp \left(-\frac{E_{j_1^{\prime}}}{k_{\mathrm{B}} T}\right)}\,, 
\end{equation}
with $\left(2 j_1+1\right)$ being the degeneracy of the rotational levels of \ch{CO}.

Thermalized rate coefficients are commonly used in radiative transfer models \cite{zoltowski2023excitation} to address collisional excitation of the target molecule, under the assumption that the rotational temperature of the projectile is equal to the kinetic temperature of the surrounding gas. Therefore, this dataset can be applied to model the abundance of \ch{HCN} in cometary comae thermalized with respect to \ch{CO}. 

To assess the accuracy of the SACM method, we compared a subset of rate coefficients with the results obtained with full quantum CC calculations.
Due to the deep potential well of the system, a direct comparison of the inelastic cross sections could indeed be biased by the presence of resonances extending over a wide range of kinetic energies. To minimize this effect, instead of the inelastic cross sections, we directly compared the state-to-state rate coefficients, computed in an energy interval between 3 and 200\,cm$^{-1}$. The 3–20\,cm$^{-1}$ range was sampled with very fine steps (0.2\,cm$^{-1}$), which gradually increased up to 2\,cm$^{-1}$ above 100\,cm$^{-1}$. To optimize the rotational basis for both \ch{HCN} and \ch{CO}, we evaluated how the inclusion of different rotational levels affected the convergence of the inelastic cross sections and we selected the basis sets that ensured convergence within 2\%. Specifically, from $2$ and $50$\,cm$^{-1}$, we included the first 22 rotational states of \ch{HCN} and the first 19 states of \ch{CO}. These were progressively increased up to 28 and 26, respectively, at $200$\,cm$^{-1}$. The angular part of the nuclear Schrödinger equation was restricted to a single partial wave $J_{\text{tot}}=0$ to keep the computational cost affordable in the cosidered energy range. The radial part was subsequently included by numerical propagation over the $R=2.9-42.0$\,\AA\, range. 
We computed CC rate coefficients in the $5-50$\,K temperature range and accounting for (de)-excitations among the rotational levels of \ch{HCN}-\ch{CO} below 80\,cm$^{-1}$. These were compared with their counterparts computed within the SACM method, using the same rotational basis used for data production but accounting only for the adiabatic channels up to $200$\,cm$^{-1}$ and for $J_{\text{tot}}=0$. The comparison for $T=10, 30$ and 50\,K is reported in Fig.\,\ref{fig3}. 

At these temperatures, most of the rate coefficients are reproduced within a factor of 2, and deviations do not exceed one order of magnitude, confirming the method's reliability. This was somewhat expected, as the agreement is consistent with the findings of \citet{godard2025promising} for the \ch{CS} and \ch{CO} system. 
\begin{figure}
\begin{center}  \includegraphics[scale=0.28]{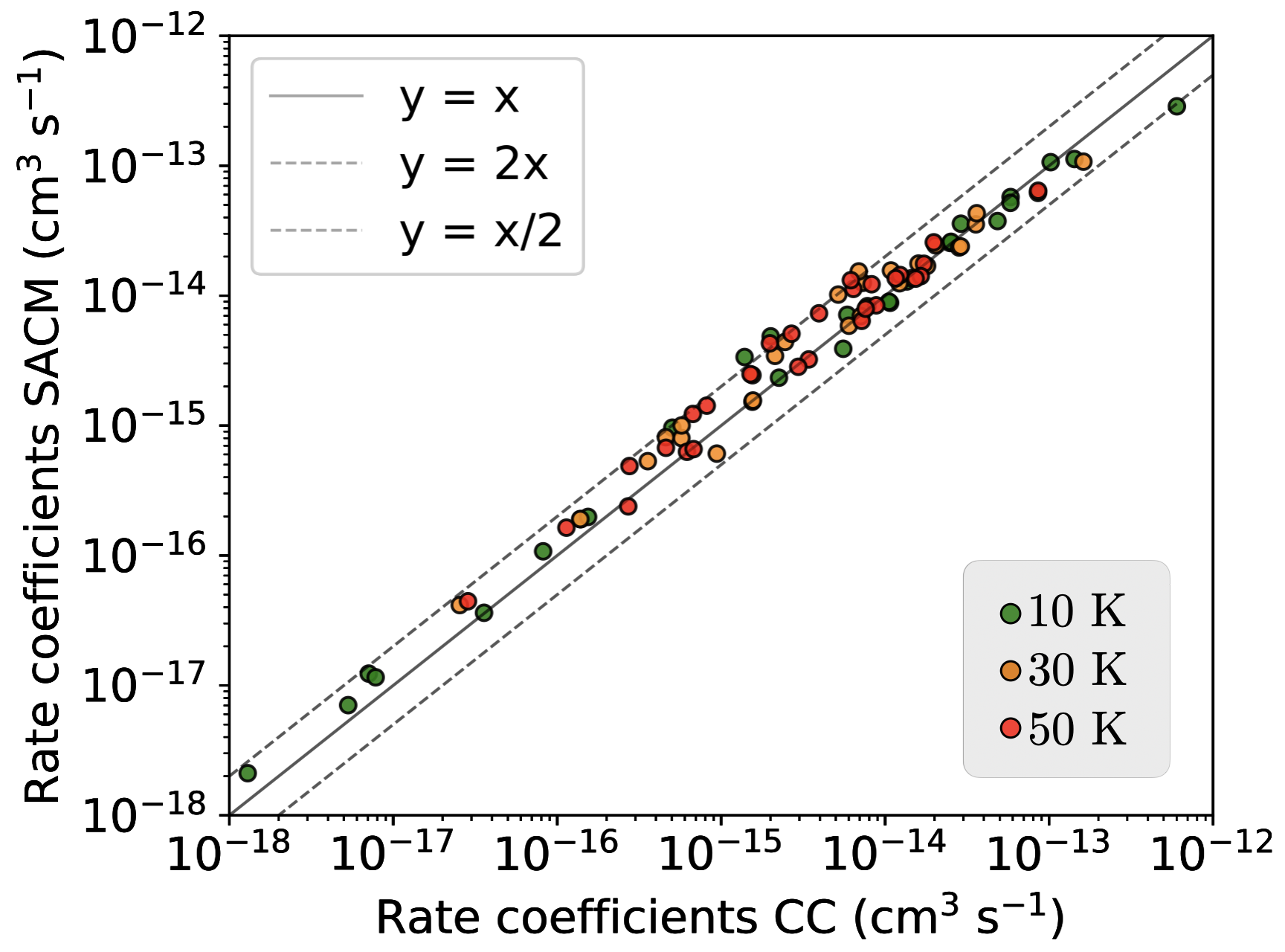}
\end{center}
\caption{Comparison between the thermalized rate coefficients for the (de)-excitation among the rotational levels of the \ch{HCN} and \ch{CO} system computed with the CC and SACM approaches at $J_{\text{tot}}=0$ for 10\,K, 30\,K and 50\,K.}
\label{fig4}
\end{figure}

The agreement between the CC and SACM approaches significantly improves when comparing thermalized rate coefficients, showing deviations of less than a factor of 2, as illustrated in Fig.\,\ref{fig4}. 
Thermal averaging over the rotational states of \ch{CO} effectively smooths out the discrepancies between the two computational methods, making the SACM approach particularly accurate in capturing the overall collisional behavior of the system.  

\section{Results and Discussion}\label{sec:disc}
\begin{figure*}
\begin{center}  \includegraphics[scale=0.41]{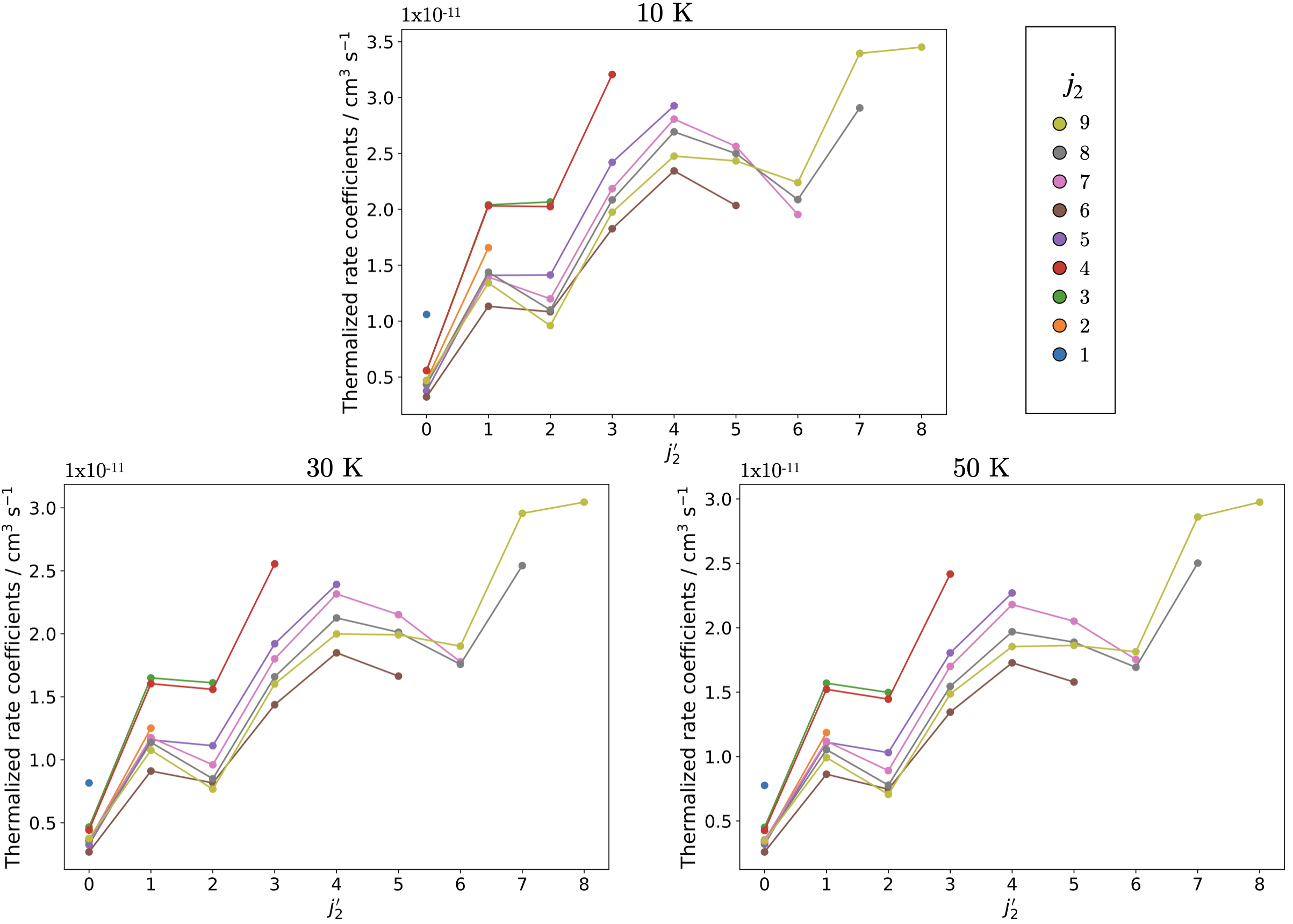}
\end{center}
\caption{Variation of the thermalized de-excitation rate coefficients as a function of $j_2'$, starting from $j_2=0-9$ and for T\,$= 10$\,K, 30\,K and 50\,K.}
\label{fig5}
\end{figure*}
\begin{figure*}
\begin{center}  \includegraphics[scale=0.49]{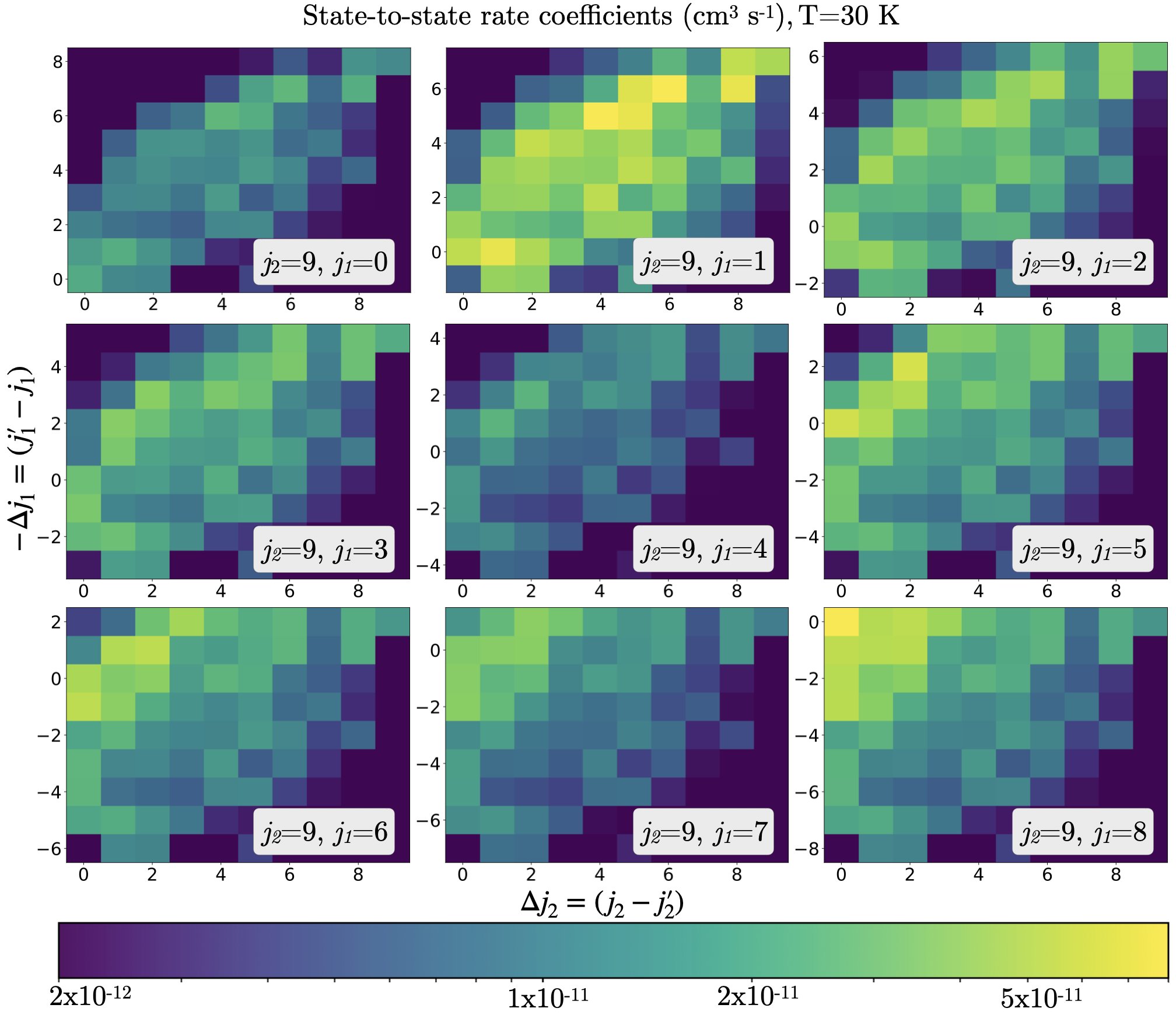}
\end{center}
\caption{Variation of some state-to-state rate coefficients as a function of $\Delta j_2$ (x axis) and $-\Delta j_1$ (y axis), starting from $j_2=9$ and $j_1=0-8$ and for T$= 30$\,K.}
\label{fig6}
\end{figure*}
We derived the first dataset of (de)-excitation rate coefficients between the lowest ten rotational levels of \ch{HCN} in collision with the first nine rotational levels of \ch{CO}. While the full dataset will be made available through the EMAA\footnote{\url{https://emaa.osug.fr//}}, LAMDA\cite{schoier2005atomic} and BASECOL\cite{dubernet2024basecol2023} databases, we summarize here the main outcomes and trends observed.

Fig.\,\ref{fig5} shows the thermalized rate coefficients of de-excitation from the first ten rotational levels of \ch{HCN}. For each initial $j_2$ level (indicated by a different color), the rate coefficients are plotted as a function of the $j_2'$ final state.
The weak dependence of these trends with respect to the kinetic temperature stands out: although absolute values of the rate coefficients are slightly higher at low temperature, the relative tendency remains largely consistent across the different thermal contributions. The same behavior (although not shown here) is observed for the state-to-state rate coefficients resolved for the rotational levels of the \ch{CO} projectile. In general, the thermalized rate coefficients tend to decrease with increasing $\Delta j_2$, reflecting the growing energy gap that they address. However, exceptions to this trend emerge when the $j_2'$ final state is equal to $1$ or $4$\,. In these cases, regardless of the initial state, a slight increase of the rate coefficient value is observed, suggesting a favor for these particular transitions. At present, the authors have not identified a clear physical rationale for this behavior.

We now address the influence of the rotational state of the \ch{CO} projectile, which is illustrated in Fig.\,\ref{fig6}. Here, the values of the state-to-state rate coefficients for de-excitations from the $j_2=9$ rotational state of \ch{HCN} and for the transitions involving the lowest nine $j_1$ rotational states of \ch{CO} are reported. The values are mapped with respect to $\Delta j_2$ (x-axis) and $-\Delta j_1$ (y-axis). Hence, the diagonal elements correspond to transitions where a de-excitation of \ch{HCN} is matched by an excitation of \ch{CO} of the same magnitude. Likewise, the elements of the first superdiagonal and subdiagonal represent transitions where the energetic jump between the levels of the two molecules differs by $\pm 1$, and so on. The largest values of the state-to-state rate coefficients are predominantly concentrated in this region, particularly for transition where $\Delta j_2=-\Delta j_1,\,\, \Delta j_2=-\Delta j_1\pm1,$ and $ \Delta j_2=-\Delta j_1\pm2$\,. This behavior is a signature of near resonant energy transfer regimes \cite{santos2011quantum,lique2012spin,lanza2014near}, where the system tends to conserve its total internal energy during the collision. This phenomenon is observed when the colliding partners, such as \ch{HCN} and \ch{CO}, have similar rotational constants, allowing their energy levels to align closely in resonance. 
Noteworthy, the near resonant energy transfer is efficiently captured by the statistical SACM method, highlighting its robustness in describing even purely quantum effects.

We expect these results to be relevant to model \ch{HCN} in cometary comae, where non-LTE radiative transfer is key for determining molecular abundances. As demonstrated in the previous section, the SACM approach provides sufficiently accurate rate coefficients, proving that the deep potential well and dense rotational structure of the system allow for a statistically valid redistribution of its energy after the formation of the intermediate complex. 
 
\section{Conclusions}\label{sec:concl}
The present study aims to support the interpretation of current and future observations of \ch{HCN} in cometary comae. In these environments, \ch{HCN} is largely abundant and widely spread, making it an important tracer of the thermal and chemical history of precometary material. Furthermore, \ch{HCN} is considered a fundamental building block in prebiotic chemistry, and its presence in comets offers valuable insights into the processes that may have contributed to the origin of life in the primordial Earth. To contribute to this understanding, we have computed the first dataset of inelastic rate coefficients involving the ten lowest rotational levels of \ch{HCN} in collision with the lowest nine states of \ch{CO}, over a temperature range of 5--50~K. 

We characterized the interaction potential using coupled-cluster electronic structure theory and an accurate fitting method. Subsequently, we performed scattering calculations by means of the SACM
statistical approach. In order to facilitate direct application in non-LTE radiative transfer models, the computed state-to-state rate coefficients were thermalized over the rotational states of \ch{CO}, assuming equilibrium between the rotational population of \ch{CO} and the kinetic temperature. 

The accuracy of the SACM method was assessed by comparison with a subset of full quantum CC calculations at $J_{\text{tot}}=0$. Given the complexity of the system and the lack of existing data, the SACM showed an encouraging agreement, reproducing state-to-state rate coefficients within a factor of 10 and thermalized rate coefficients within a factor of 2. 

With the contribution of this work, accurate collisional datasets for \ch{HCN} interacting with both \ch{H2O}\cite{dubernet2019first,zoltowski2025collisional} and \ch{CO} are now available. Given the ubiquity of \ch{HCN} in cometary comae, these datasets offer a valuable test case to investigate the sensitivity of non-LTE effects under diverse physical conditions, such as different heliocentric distances. Worth to note, also \ch{CO2} is quite abundant in many cometary comae, especially at intermediate heliocentric distances. However, its high density of rotational states (resulting from a very low rotational constant) makes its collisional characterization significantly more demanding from a computational point of view. 
Although it is not expected to be the dominant collisional partner, computing rate coefficients with \ch{CO2} could nonetheless provide additional insights.
All of these data will play a central role in driving the non-LTE distribution of population of \ch{HCN} in cometary comae, bringing new knowledge to probe the chemical and physical conditions of the early Solar System. 


\begin{acknowledgement}
This work has been supported by Region Bretagne. The authors acknowledge the ``Programme National de Planétologie" (PNP) under the responsibility of INSU, CNRS (France). FL acknowledges the Institut Universitaire de France.
RD, EQS, and ABP are supported by the U.S. Department of Energy under Award DE-SC0025420.

\end{acknowledgement}

\bibliography{bibliography}

\section{TOC Graphic}

\begin{figure}
    \centering
    \includegraphics[width=8.25cm]{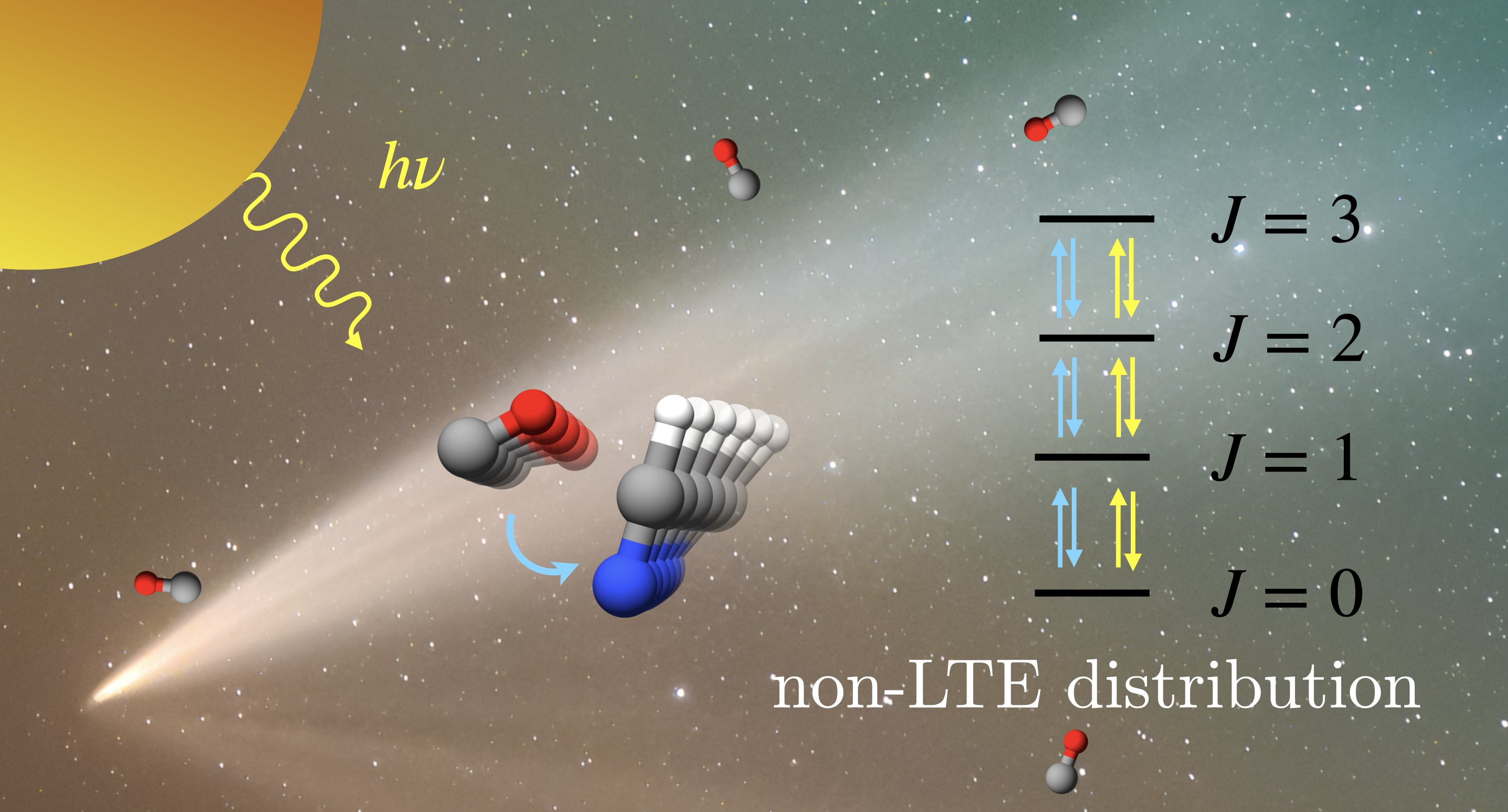}
\end{figure}

\end{document}